\documentclass[11pt,a4paper]{article}

\pdfoutput=1
\usepackage{jheppub}
\usepackage{amsfonts}
\usepackage{amsmath}
\usepackage{amssymb}
\usepackage{graphicx,color}
\usepackage{float}
\usepackage{hyperref}
\usepackage{subfigure}

\title{Adiabatic out-of-equilibrium solutions to the Boltzmann equation in warm inflation}

\author[a]{Mar Bastero-Gil}    
\author[b]{Arjun Berera}    
\author[c]{Rudnei O. Ramos}    
\author[d]{Jo\~{a}o G.~Rosa} 

\affiliation[a]{Departamento de F\'{\i}sica Te\'orica y del Cosmos, Universidad de Granada, Granada-18071, Spain}
\affiliation[b]{School of Physics and Astronomy, University of Edinburgh, Edinburgh, EH9 3FD, United Kingdom}
\affiliation[c]{Departamento de F\'{\i}sica Te\'orica, Universidade do Estado do Rio de Janeiro, 20550-013 Rio de Janeiro, RJ, Brazil}
\affiliation[d]{Departamento de F\'{\i}sica da Universidade de Aveiro and CIDMA,  Campus de Santiago, 3810-183 Aveiro, Portugal}

\emailAdd{mbg@ugr.es}
\emailAdd{ab@ph.ed.ac.uk}
\emailAdd{rudnei@uerj.br}
\emailAdd{joao.rosa@ua.pt}

\abstract{
We show that, in warm inflation, the nearly constant Hubble rate and
temperature lead to an adiabatic evolution of the number density of
particles interacting with the thermal bath, even if thermal
equilibrium cannot be maintained. In this case, the number density is
suppressed compared to the equilibrium value but the associated
phase-space distribution retains approximately an equilibrium form, with a smaller amplitude and a slightly smaller effective temperature. As
an application, we explicitly construct a baryogenesis mechanism
during warm inflation based on the out-of-equilibrium decay of
particles in such an adiabatically evolving state. We show that this
generically leads to small baryon isocurvature perturbations, within
the bounds set by the Planck satellite. These are correlated with the
main adiabatic curvature perturbations but exhibit a distinct spectral
index, which may constitute a smoking gun for baryogenesis during warm
inflation. Finally, we discuss the prospects for other applications of adiabatically evolving out-of-equilibrium states.}


\arxivnumber{1711.09023 [hep-ph]}

\begin{document}
\maketitle

%%%%%%%%%%%%%%%%%%%%%%%%%%%%%%%%%%%%%%%%%%%%%%%%%%%%%%%%%%%%%%%%
\section{Introduction}

Equilibrium and out-of-equilibrium dynamics play central roles in
cosmology, being crucial in determining, for instance, the present
abundances of light nuclei, dark matter and other thermal relics or
the cosmological baryon asymmetry. {}From the nearly perfect blackbody
spectrum of the Cosmic Microwave Background (CMB) radiation and the
successful predictions of Big Bang Nucleosynthesis (BBN) for the
abundances of Helium and other light nuclei, we can infer that the
Universe achieved a state very close to thermal equilibrium in the
early stages of its evolution. Most of the cosmological dynamics is
based on determining when a given particle species was in equilibrium
with the thermal cosmic plasma and when it decoupled from the
latter. This generically involves solving intricate systems of coupled
Boltzmann equations for different particle species, but it typically
suffices to use the well-known ``rule of thumb" that a particle is in
equilibrium with the cosmic plasma whenever its interaction rate with
the latter exceeds the Hubble expansion rate, $\Gamma \gtrsim
H$. Since in most cosmological eras the Hubble rate varies
significantly in a Hubble time, this implies that the transition from
an equilibrium to an out-of-equilibrium (decoupled) state is very fast
on cosmological time scales.

An exception to this rule is naturally the period of
inflation~\cite{inflation}, where the Hubble rate remains nearly
constant for about 50-60 e-folds of expansion required to solve the
flatness and horizon problems of the Big Bang model. In canonical
models of inflation the question of whether or not a particle species
is in thermal equilibrium makes little sense, since the exponentially
fast expansion quickly dilutes away all particle species present in
the pre-inflationary epochs. However, in the context of warm
inflation~\cite{Berera:1995wh, Berera:1995ie} this question plays a
prominent role. In this alternative paradigm, dissipative processes
continuously transfer the inflaton's energy into the cosmic plasma,
leading to particle production that sustains a slowly varying
temperature during inflation. There are several reasons to consider
such an alternative paradigm. In such scenarios,
dissipative friction damps the inflaton's evolution and prolongs inflation, with
radiation smoothly taking over as the dominant component if strong
dissipation is attained at the end of the slow-roll evolution regime
(see, e.g.,~Refs.~\cite{Berera:2008ar, BasteroGil:2009ec,
  Berera:1996fm, Berera:1999ws}). The greatest appeal of warm
inflation lies perhaps in the fact that thermal inflaton fluctuations
are directly sourced by dissipative processes, changing the form of
the primordial spectrum of curvature perturbations and thus providing a
unique observational window into the particle physics behind
inflation~\cite{Berera:1999ws, Hall:2003zp, Moss:2007cv,
  Graham:2009bf, Ramos:2013nsa, Bartrum:2013oka,
  Bartrum:2013fia,Bastero-Gil:2014raa}. In addition, we have recently
shown that warm inflation can be consistently realized in a simple
quantum field theory framework requiring very few fields, 
the {\it Warm Little Inflaton}
scenario~\cite{Bastero-Gil:2016qru}, where the required flatness of
the inflaton potential is not spoiled by thermal effects (see also
Refs.~\cite{Berera:1998px, Berera:2002sp, Moss:2006gt,
  BasteroGil:2010pb, BasteroGil:2011mr, BasteroGil:2012cm,
  Cerezo:2012ub} for earlier alternative models), paving the way for
developing a complete particle physics description of inflation that
can be fully tested with CMB and Large-Scale Structure (LSS)
observations and possibly have implications for collider
and particle physics data.

Independently of the particle physics involved in sustaining a thermal
bath during inflation, a generic feature of warm inflation is the slow
evolution of both the temperature and the Hubble parameter for the
usually required 50-60 e-folds of expansion. Since both scattering and
particle decay rates typically depend on the former, this implies that
the ratio $\Gamma/H$ will generically evolve slowly during
inflation. Consequently, as we explicitly show in this work for the
first time, particle species in the warm inflationary plasma can
maintain distributions that are slowly evolving and out-of-equilibrium  
throughout
inflation, whether or not they are directly involved in the
dissipative dynamics. This may have an important impact not only on
the inflationary dynamics and predictions themselves but also on the
present abundance of different components. As an example of
application of this novel observation, we show that this can lead to
the production of a baryon asymmetry during inflation, a possibility
that can be tested in the near future with CMB and LSS observations.

This work is organized as follows. In Sec~\ref{sec2} we analyze the
Boltzmann equation for a particle species interacting with a thermal
bath for an adiabatic evolution of the ratio $\Gamma/H$, focusing
explicitly on the case of decays and inverse decays for
concreteness. We then show that this leads to slowly varying
out-of-equilibrium configurations, obtaining the overall particle
number density and its phase space distribution. We briefly discuss how similar results can be obtained for scattering processes. In Sec.~\ref{sec3} we
use these results to develop a generic baryogenesis (or leptogenesis)
mechanism during inflation. Then, in Sec.~\ref{sec4}, we discuss how
this generically leads to baryon isocurvature modes that give a small
contribution to the primordial curvature perturbation spectrum. In
Sec.~\ref{sec5} we summarize our results and discuss their potential
impact on other aspects of the inflationary and post-inflationary
history. An Appendix is included where some technical details are
given.

%%%%%%%%%%%%%%%%%%%%%%%%%%%%%%%%%%%%%%%%%%%%%%%%%%%%%%%%%%%%%%%%
\section{Adiabatic out-of-equilibrium dynamics}
\label{sec2}

Let us consider a particle $X$ interacting with a thermal bath at
temperature $T$ in an expanding  Universe\footnote{Throughout this
  work we consider  a spatially flat
  {}Friedmann-Lemaitre-Robertson-Walker (FLRW) space-time in which the
  metric is given by $ds^2  =  dt^2 - a(t)^2 d{\bf x}^2$, where
  $t$ is physical time, ${\bf x}$ are the comoving spatial coordinates
  and $a(t)$ is the cosmological scale factor.}. For concreteness, let us explicitly consider the case where $X$
decays into particles $Y_1$ and $Y_2$ in the thermal bath, and assume
that these maintain a near equilibrium distribution through other processes
that do not directly involve $X$. Let us also, for simplicity, focus
on the case where $X$ is a boson with $g_X$ degrees of freedom and  the decaying products $Y_1$ and
$Y_2$ are either bosons or fermions  (extending our results to
decaying fermions is straightforward).
The number density of $X$ particles, 
\begin{equation}
  n_X=g_X \int \frac{d^3p}{(2\pi)^3} f_X(\mathbf{p})~,
\end{equation} 
has an evolution, in an expanding flat FLRW universe, that is
described by the  Boltzmann
equation~\cite{Kolb:1990vq,lyth2009primordial}
\begin{eqnarray} \label{boltzmann_n}
\dot{n}_X+3H n_X= C~,
\end{eqnarray}
where the collision term $C$ includes the effects of decays
$X\rightarrow Y_1+Y_2$ and inverse decays $Y_1+Y_2\rightarrow X$ and
takes the form\footnote{The inclusion of Landau damping effects will
  not change our conclusions.}
\begin{eqnarray}
C=\! -\!\!\int\!\! d\Pi_X d\Pi_{1}d\Pi_{2}
(2\pi)^4\delta^4(p_X-p_1-p_2)|\mathcal{M}|^2
\left[f_X (1 \pm f_1^{eq})(1 \pm f_2^{eq})-f_1^{eq} f_2^{eq}
  (1+f_X)\right],\nonumber\\
\label{coll}
\end{eqnarray}
where $d\Pi_i =g_i d^3 p_i/2E_i (2\pi)^3$, $\mathcal{M}$ is the matrix element associated with both the decay and
inverse decay processes (related by CPT invariance) and $f_i$ are
the phase-space distribution functions for each particle.  Note also
that the plus (minus) sign in $1 \pm f_i$ in Eq.~(\ref{coll})
refers to bosons (fermions) and it is related to the usual Bose
enhancement (Pauli-blocking) effect.  Since we assume
$Y_{1,2}$ are in equilibrium, we have
\begin{eqnarray}
f_{i}^{eq}={1\over e^{\beta(E_{i}-\mu_{i})}\pm 1}~,\qquad i=1,2
\end{eqnarray}
where $\beta=1/T$ in natural units and the plus (minus) sign is for a
Fermi-Dirac (Bose-Einstein) distribution.  Taking into account conservation of energy, $E_X=E_1+E_2$, and
assuming chemical equilibrium,  $\mu_X=\mu_1+\mu_2$ (an assumption
that we may drop if chemical potentials can be neglected), then, it is
easy to show that
\begin{eqnarray}
1 \pm f_1^{eq} \pm f_2^{eq} = {f_1^{eq} f_2^{eq} \over f_X^{eq}}~,
\end{eqnarray}
where
\begin{eqnarray}
 f_X^{eq}={1\over e^{\beta(E_{X}-\mu_{X)}}-1}
\end{eqnarray}
is the distribution of the $X$ particles when they are in
equilibrium. This then allows us to write the collision term in the
form
\begin{eqnarray}
C&=& -\int d\Pi_X d\Pi_{1}d\Pi_{2}
(2\pi)^4\delta^4(p_X-p_1-p_2)|\mathcal{M}|^2
{f_1^{eq} f_2^{eq} \over f_X^{eq}}\left( f_X-
f_X^{eq}\right)\nonumber\\ &=& - \int {d^3p_X\over (2\pi)^3}\Gamma_X
(f_X- f_X^{eq})~,
\end{eqnarray}
where the equilibrium decay width of the $X$ boson is given by
\begin{eqnarray}
\Gamma_X&= &{1\over 2E_X f_X^{eq}}
\int\!\! d\Pi_{1}d\Pi_{2}
(2\pi)^4\delta^4(p_X-p_1-p_2)|\mathcal{M}|^2 f_1^{eq}f_2^{eq}~.\nonumber
\end{eqnarray}
We can simplify further the Boltzmann equation by discarding the
momentum-dependence of the decay width, a common procedure in the
literature, by considering its thermal average~\cite{Kolb:1990vq}:
\begin{eqnarray}
\bar\Gamma_X={1\over n_X^{eq}}\int d^3p_X \Gamma_X f_X^{eq}~.
\label{barGamma}
\end{eqnarray}
The Boltzmann equation (\ref{boltzmann_n}) can then be cast into the
familiar form
\begin{eqnarray}
\dot{n}_X+3H n_X= -\bar\Gamma_X(n_X-n_X^{eq})~.
\end{eqnarray}
It is straightforward to obtain an analogous result for an arbitrary
number of particles in the final state.

In a non-expanding Universe, the collision term will then naturally
drive the number density of $X$ particles towards its equilibrium
value, while in an expanding Universe this only occurs for
$\bar\Gamma_X\gg H$, which is the familiar rule of thumb in
cosmology. Now, in a cosmological setting such as warm inflation,
since $H$ and $T$ are slowly-varying and $\bar\Gamma_X$ will depend
only on $T$ and on the masses of the parent and daughter particles, we
can take the ratio $\bar\Gamma_X/H$ to be slowly-evolving.  {}For all
cases where $\gamma_X \equiv \bar\Gamma_X/H$ is slowly-evolving on the
Hubble scale, i.e.,~$\dot\gamma_X/\gamma_X \ll H^{-1}$, we may take
$\gamma_X$ to be a constant as a first approximation and write the
Boltzmann equation in terms of the number of e-folds of expansion,
$dN_e= H dt$, as
\begin{eqnarray} \label{Boltzmann}
n_X'+3 n_X= -\gamma_X(n_X-n_X^{eq})~.
\end{eqnarray}
If the temperature is slowly-varying, we may take $n_X^{eq}$ to be
constant as well, yielding the solution
\begin{eqnarray} \label{adiabatic_sol}
n_X(N_e)= {\gamma_X\over 3+\gamma_X} n_X^{eq}\left(1-e^{-(3+\gamma_X)N_e}\right) + n_X(0)
e^{-(3+\gamma_X) N_e}~.
\end{eqnarray}
Hence, $X$ is driven exponentially fast to the solution
\begin{eqnarray} \label{stationary_sol}
n_X\simeq {\gamma_X\over 3+\gamma_X} n_X^{eq} ~.
\end{eqnarray}
Note that this is not exactly stationary due to the slow variation of
both $\gamma_X$ and $n_X^{eq}$, so we refer to this solution as {\it
  adiabatic}. This solution is attained in less than a e-fold if the
initial number density is not too far from the quasi-stationary value,
and even large discrepancies will be quickly washed away within a few
e-folds. This agrees with the statement made above that $n_X$ is
driven towards its equilibrium value for $\gamma_X\gg 1$, but reveals
a novel feature that is absent in general cosmological settings
{\textemdash} that even for $\gamma_X\ll 1$ a small but not
necessarily negligible number density of $X$ particles remains
constant despite the fast expansion of the Universe.  This is thus a
new type of solution that only arises in cosmological settings where
the temperature and Hubble rate are varying  
slowly, such as warm inflation. In an adiabatic approximation, we
can include the slow 
variation of these parameters, namely $\gamma_X=\gamma_X(N_e)$ and
$n_X^{eq}=n_X^{eq}(N_e)$ in the quasi-stationary solution in
Eq.~(\ref{stationary_sol}).

\begin{figure}[!htb]
\begin{center}
\subfigure[Decay into bosons.]  {
  \includegraphics[width=0.6\textwidth]{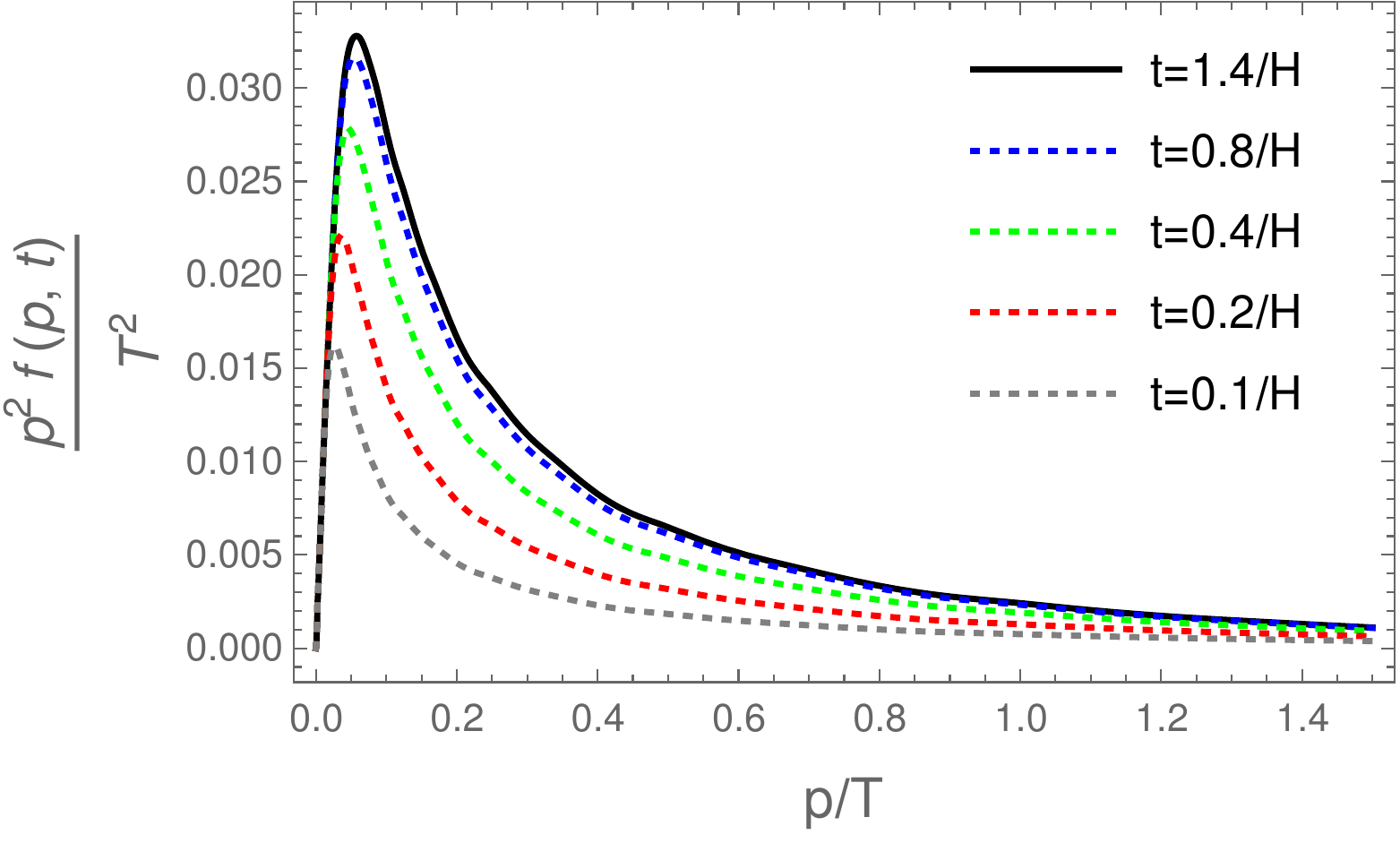}}
\subfigure[Decay into fermions.]{
  \includegraphics[width=0.6\textwidth]{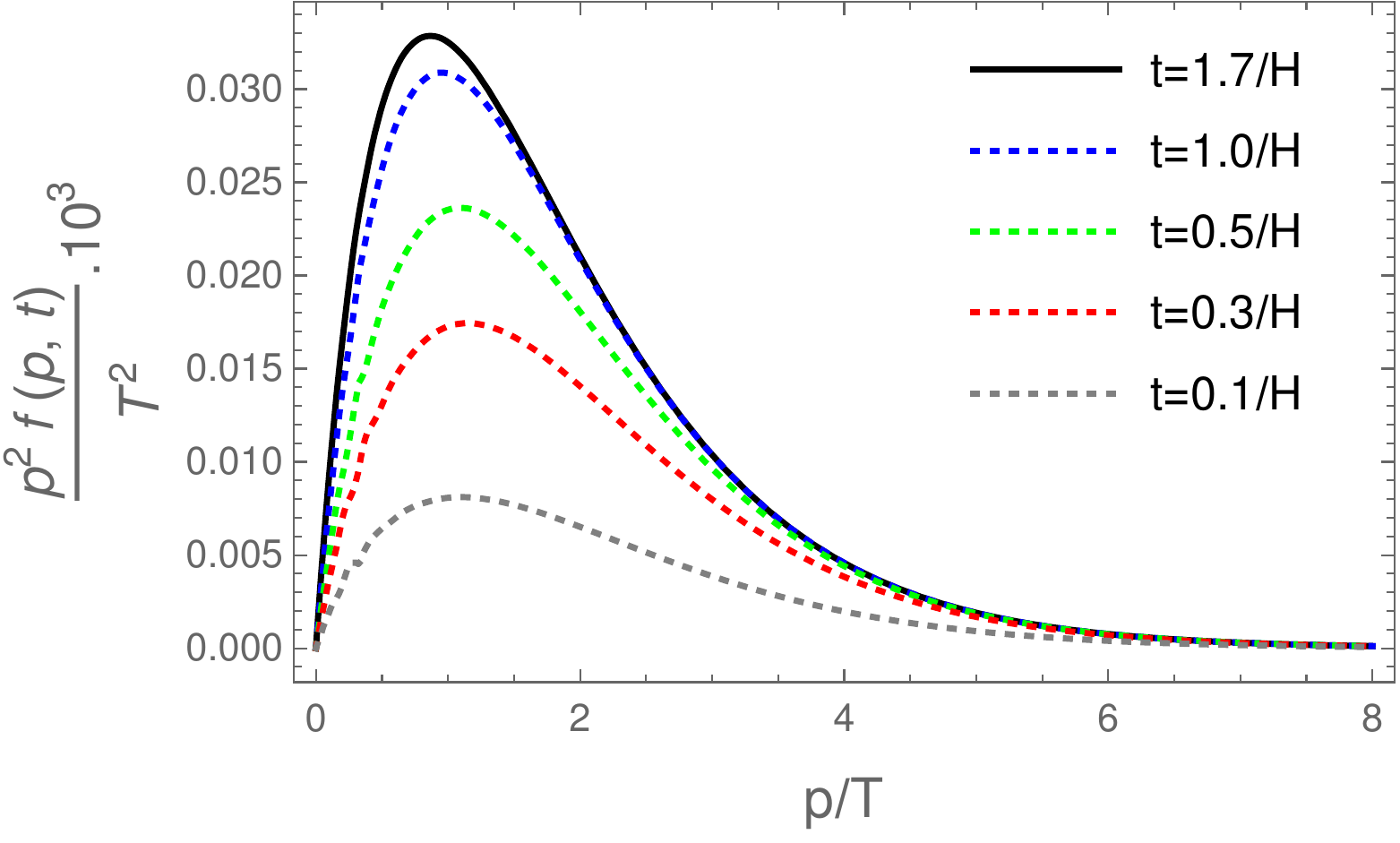}}
\end{center}
\caption{The phase space distribution at different times (in units of the Hubble parameter), as a function of the momentum
  $p$, for decay into bosons (a) and for decay into fermions (b). The
  parameters used are $m_X=0.001T$,  and $\Gamma_0^{(B,F)}/H=0.5$,
  which corresponds to $\gamma_X^{(B)} =0.0245$ (bosons) and $\gamma_X^{(F)} =10^{-4}$ (fermions).}
\label{fig1} 
\end{figure}

We may also take a step further and compute the phase-space
distribution of $X$-particles, which follows the momentum-dependent
Boltzmann equation in a flat FLRW universe,
\begin{equation} 
{\partial f_X({\bf p},t)\over \partial t} - H p  {\partial f_X({\bf
    p},t)\over \partial p} = C({\bf p})~,
\label{boltzmann}
\end{equation}
where the collision term is given by 
\begin{equation}
C({\bf p})= -\Gamma_X({\bf p}) \left[ f_X({\bf p},t) - f_X^{ eq}({\bf
    p}) \right],
\end{equation} 
from the results obtained above. For illustrative purposes, we will
consider two distinct cases where an $X$ scalar boson decays into
either fermion or scalar boson pairs, through Yukawa or
scalar trilinear interactions, respectively. We neglect all chemical
potentials for simplicity. The corresponding decay widths are given
by~\cite{BasteroGil:2010pb}
\begin{eqnarray} \label{boson_fermion_widths}
\Gamma_X^{(B)}\!\!&=&\!\! \Gamma_{0}^{(B)}{m_X\over \omega_p} \left[1+
  2{T\over
    p}\log\left({1-e^{-\omega_+/T}\over1-e^{-\omega_-/T}}\right)\right]
\nonumber\\ \Gamma_X^{(F)}\!\!&=&\!\!
\Gamma_{0}^{(F)}{m_X\over \omega_p}  \left[1+ 2{T\over
    p}\log\left({1+e^{-\omega_+/T}\over1+e^{-\omega_-/T}}\right)\right]
\end{eqnarray}
where we have neglected the masses of the decay products, with
$\omega_{\pm} = (\omega_p\pm p)/2$, $\omega_p=\sqrt{p^2+m_X^2}$,
$p=|{\bf p}|$. The decay widths at zero-temperature and zero-momentum
are, in the two cases, given by
\begin{eqnarray} \label{boson_fermion_widths0}
\Gamma_0^{(B)}={g_B^2\over 32\pi}{M^2\over m_X}~, \qquad
\Gamma_0^{(F)}= {g_F^2 \over 8\pi} m_X ,
\end{eqnarray}
where $g_{B,F}$ are dimensionless couplings and $M$ is the mass scale
of the trilinear scalar coupling. We have solved the Boltzmann
equation~(\ref{boltzmann}) numerically in both cases for different
values of the $X$ boson mass and decay width, taking $f_X({\bf p},
0)=0$ and imposing $ f_X({\bf p}, t)\rightarrow 0$ in the limit
$p\rightarrow \infty$ (in practice at a sufficiently large momentum
value). We illustrate the resulting time evolution of the phase-space
distribution  for the bosonic and fermionic cases in {}Fig.~\ref{fig1}(a)
and {}Fig.~\ref{fig1}(b) respectively. In both cases
the decay is of a relativistic $X$ boson with mass set at the value
$m_X=0.001T$ and with $\Gamma_0^{(B,F)}/H=0.5$, which corresponds to
$\gamma_X^{(B)} =0.0245$ ($\gamma_X^{(F)} =10^{-4}$ in the case of
decay into fermions). It is clear  from the results shown in
{}Fig.~\ref{fig1} that in both cases the distribution reaches a
stationary configuration after only a couple of Hubble times, a
result that we obtained generically for different values of the mass
and decay width, for both fermionic and bosonic decay. Naturally, for
larger values of $\gamma_X$ the stationary configuration is
attained faster.

%%%%%%%%%%%%%%%%%%%%%%%%%%%%%%%%%%%%%%%%%%%%%%%%%%%%%%%%%%%%%%%%%%%%%%%
%%%%%%%%%%%%%%%%%%%%%%%%%%%%%%%%%%%%%%%%%%%%%%%%%%%%%%%%%%%%%%%%%%%%%%%%

These stationary solutions can be obtained by setting $\partial
f_X/\partial t=0$ in the Boltzmann equation (\ref{boltzmann}),
yielding a first-order inhomogeneous differential equation for
$f_X^{stat}({\bf p})$. Following standard methods, we may formally
write the stationary solution in the integral form
\begin{eqnarray} \label{integral_solution}
f_X^{stat} (p)= f_X^{(h)} (p)\int_p^{\infty} {dp'\over p'}
{\Gamma_X(p')\over H} {f_X^{eq}(p')\over f_X^{(h)}(p')}~,
\end{eqnarray}
where $f_X^{(h)}= e^{\int dp' \Gamma_X(p')/ H p'}$ is the homogeneous
solution, which takes the form $f_X^{(h)}= p^{\gamma_X}$ if one
replaces the decay width by its thermal average Eq.~(\ref{barGamma}),
as explained above. 

\begin{figure}[!htb]
\begin{center}
\subfigure[Decay into bosons.]  {
  \includegraphics[width=0.6\textwidth]{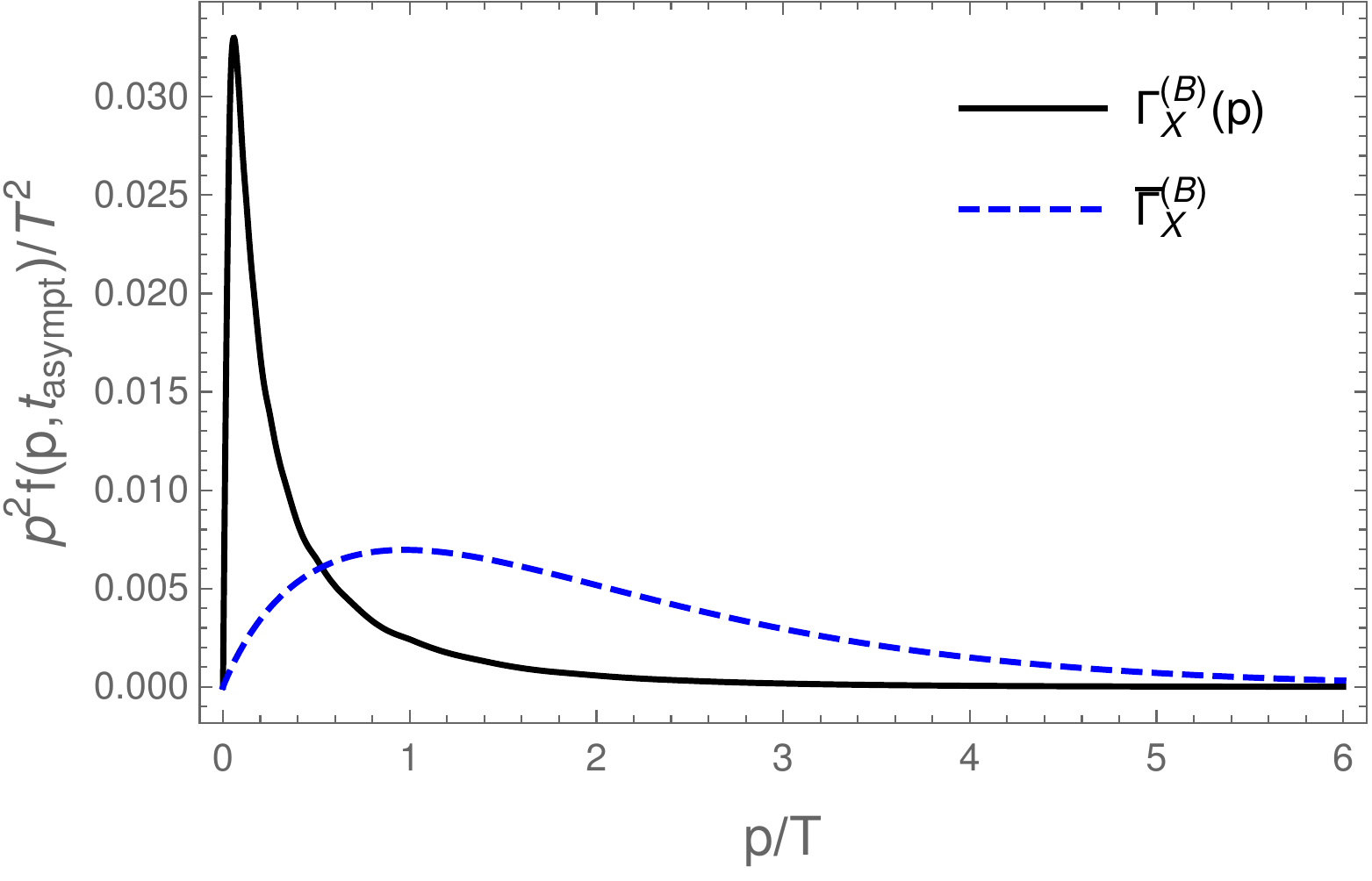}}
\subfigure[Decay into fermions.]{
  \includegraphics[width=0.6\textwidth]{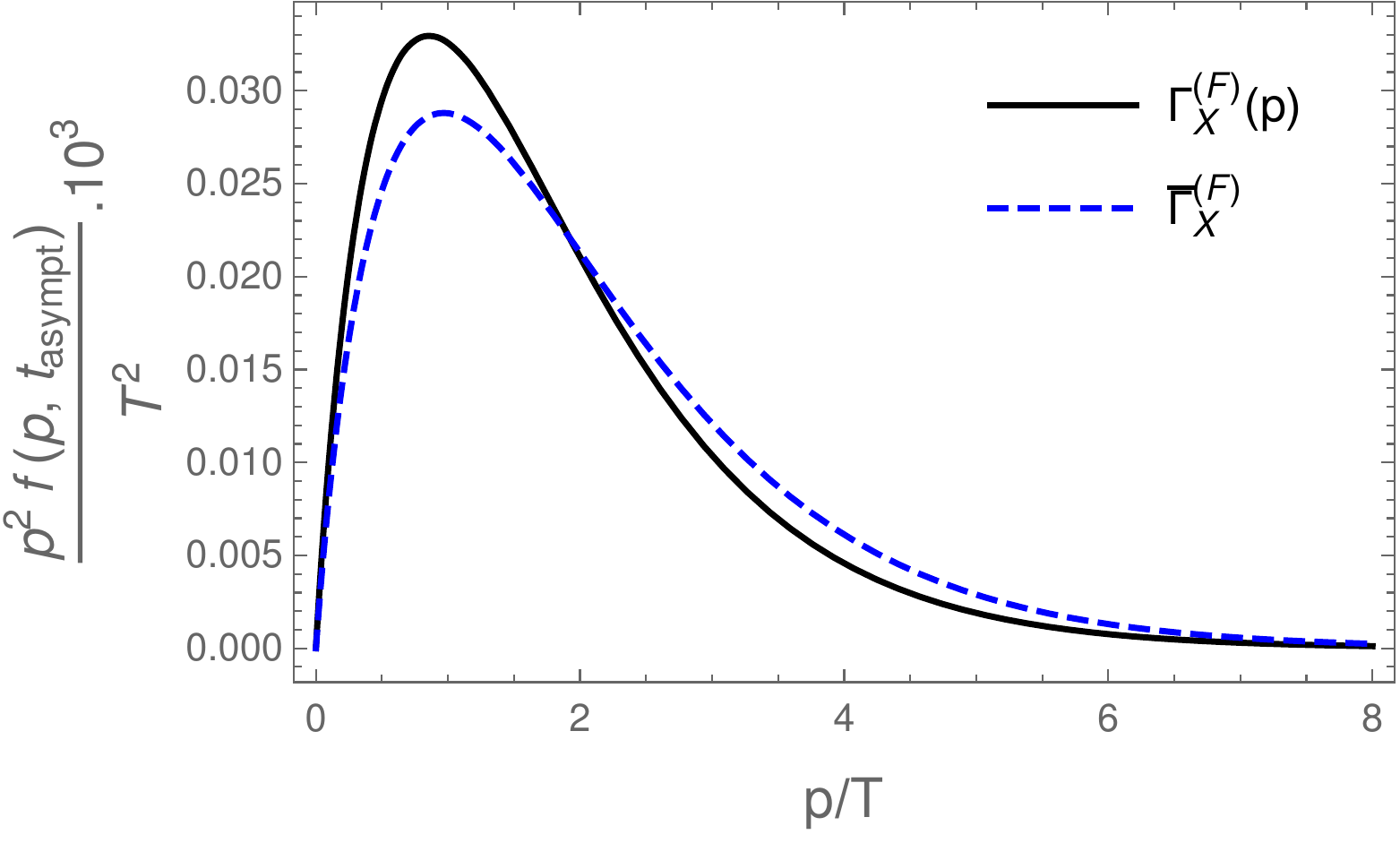}}
\end{center}
\caption{The phase space distribution, as a function of the momentum
  $p$,  at large asymptotic times (when the distributions have already
  reached the stationary state), for decay into bosons (a) and for
  decay into fermions (b). The parameters used are $m_X=0.001T$ and
  $\Gamma_0^{(B,F)}/H=0.5$, which corresponds to $\gamma_X^{(B)} =0.0245$ (bosons) and $\gamma_X^{(F)} =10^{-4}$ (fermions). Solid lines yield the solution when considering the full momentum dependent decay widths, while
  dashed lines correspond to the solution obtained using the constant thermally averaged decay widths. }
\label{fig2} 
\end{figure}

{}For practical purposes, however, this integral form is not very
useful, since integrals involving the Bose-Einstein distribution do
not have, in general, a simple analytical form and have to be computed
numerically.  In {}Fig.~\ref{fig2}, we show the obtained stationary
distributions for bosonic and fermionic decays for the same values of
the $X$ mass and average decay width considered in {}Fig.~\ref{fig1}. 

%%%%%%%%%%%%%%%%%%%%%%%%%%%%%%%%%%%%%%%%%%%%%%%%%%%%%%%%%%%%%%%%%%%%%%%
\begin{figure}[!htb]
\begin{center}
\subfigure[The fit amplitude $A$.]  {
 \includegraphics[width=0.57\textwidth]{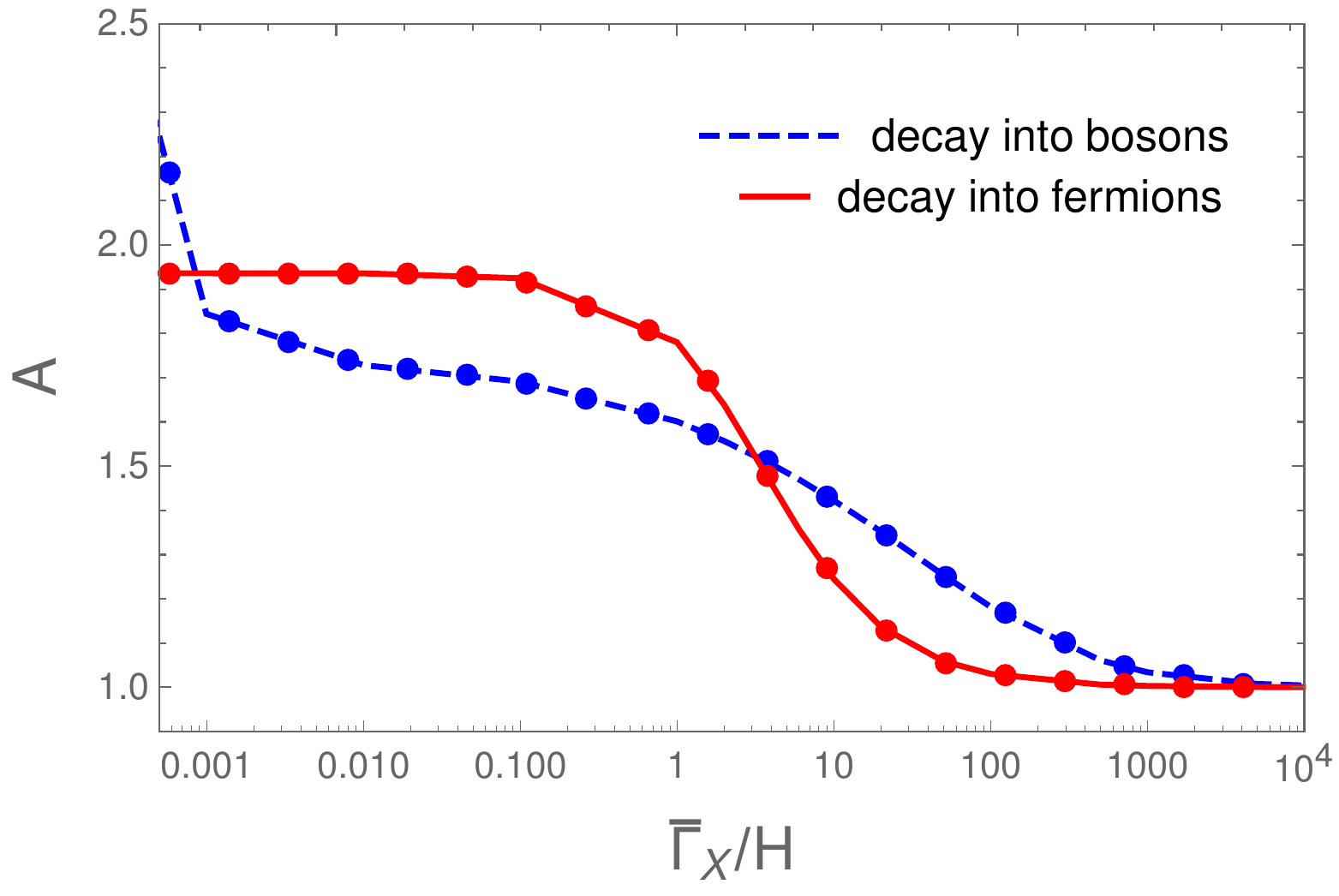}} \subfigure[The
  effective temperature $T_{\rm stat}$.]{
  \includegraphics[width=0.57\textwidth]{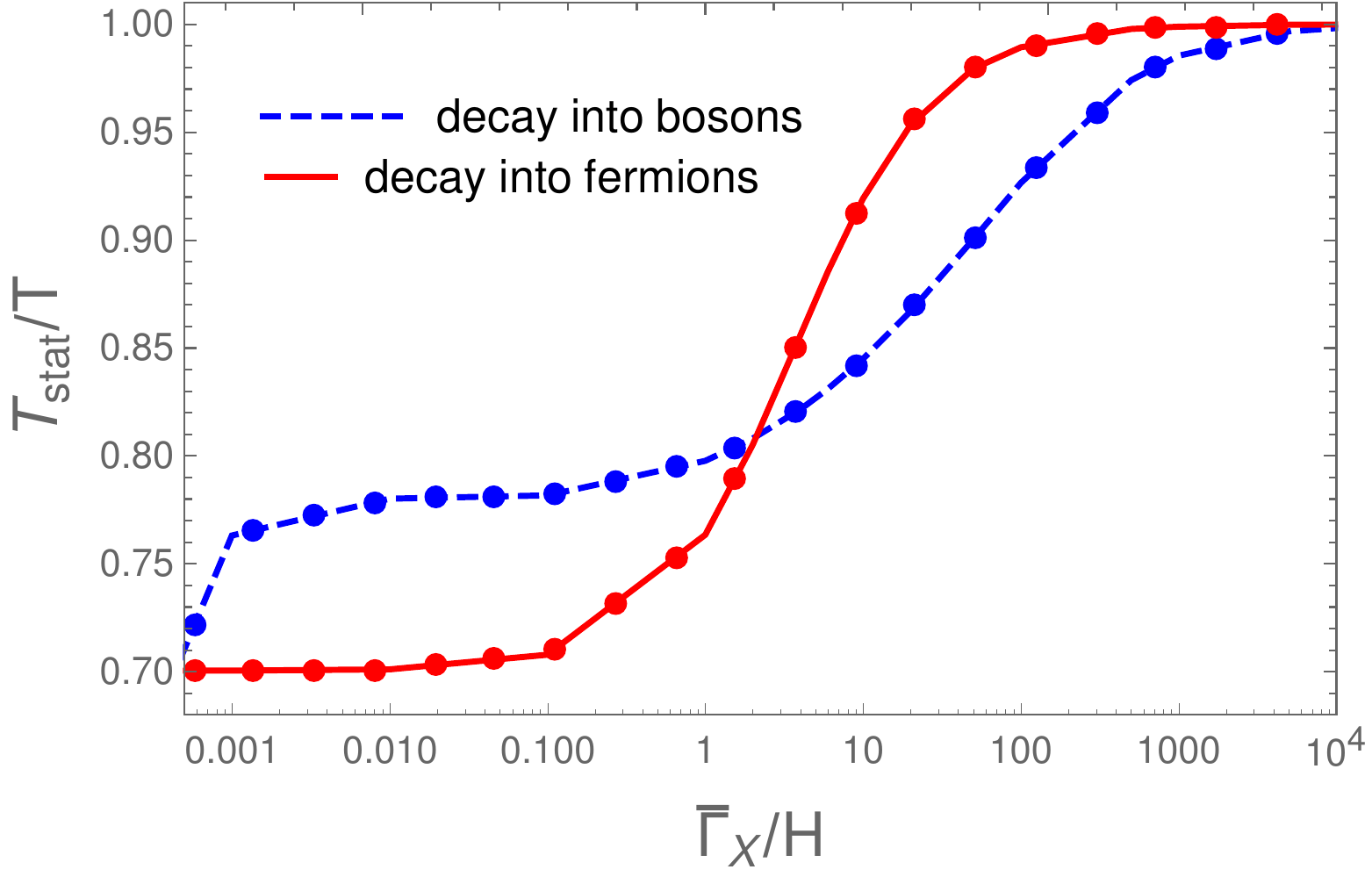}}
\end{center}
\caption{The fitting amplitude $A$ (a) and effective temperature
  $T_{\rm stat}$ (b) for the stationary distribution with $m_X=0.001T$. Dashed lines are
  for the case of decay into bosons, while solid lines are for the
  case of decay into fermions. }
\label{fig3} 
\end{figure}

%%%%%%%%%%%%%%%%%%%%%%%%%%%%%%%%%%%%%%%%%%%%%%%%%%%%%%%%%%%%%%%%%%%%%%%%
{}For comparison, we also show in {}Fig.~\ref{fig2} the distribution
obtained when replacing $\Gamma_X(p)$ by $\bar\Gamma_X$. We can see
that for fermionic decay this yields a good approximation to the full
solution, while for bosonic decay there are more prominent
differences. In particular, the latter distribution is peaked at lower
momentum values than the one obtained using $\bar\Gamma_X$, which is
related to the Bose enhancement of the decay at low-momentum values
$p\lesssim m_X$. {}For $m_X\ll T$ there is thus a substantial
variation of the bosonic decay width with momentum in the relevant
range $p\lesssim T$, while for fermionic decay it is a good
approximation to use the thermally averaged decay width in place of
the full momentum-dependent expression. Note that the larger the mass
of the $X$ boson the closer the distributions are to the one obtained
using $\bar\Gamma_X$, since thermal corrections to the decay width
become less important in this regime for $p\lesssim T \lesssim m_X$.
All stationary distributions are nevertheless reasonably well fitted
by an expression of the form
\begin{equation}
f_X^{\rm stat}(p) = A \frac{\Gamma_X (p,T_{\rm stat})}{3H +
  \Gamma_X(p,T_{\rm stat})} \frac{1}{e^{\sqrt{p^2+m_X^2}/T_{\rm
      stat}}-1}~.
\label{fit}
\end{equation}
We show in {}Fig.~\ref{fig3} the results for the fit amplitude $A$ and
effective temperature $T_{\rm stat}$  for $m_X=0.001 T$ and different
values of the thermally averaged decay width, for both the bosonic and
fermionic decays.

%%%%%%%%%%%%%%%%%%%%%%%%%%%%%%%%%%%%%%%%%%%%%%%%%%%%%%%%%%%%%%%%%%%%%%%
%%%%%%%%%%%%%%%%%%%%%%%%%%%%%%%%%%%%%%%%%%%%%%%%%%%%%%%%%%%%%%%%%%%%%%%%

As we can see in {}Fig.~\ref{fig3}, for relativistic $X$ bosons the
coefficient $A\sim \mathcal{O}(1)$ for fermionic decays and also
bosonic decays, unless the decay width is much smaller than the Hubble
parameter, while the effective temperature $T_{\rm stat}\lesssim
T$. The larger variation of the fit parameters for bosonic decay is
again due to the above mentioned Bose enhancement effect, a variation
that becomes smaller for larger values of the $X$ mass.  With the
results shown in {}Fig.~\ref{fig3}, we also note that, generically, the
effective temperature $T_{\rm stat}$ approaches (asymptotically) the
thermal bath temperature $T$ from below and likewise for the overall
amplitude in front of Eq.~(\ref{fit}). This then means that the
stationary solution $f_X^{\rm stat}$ gets suppressed at large momenta
relative to the equilibrium one $f^{eq}$. This result is similar to
the one obtained recently for an exact solution of the Boltzmann
equation in a FLRW background~\cite{Bazow:2015dha}, though it differs
fundamentally from the solution found here. In particular, the
result of Ref.~\cite{Bazow:2015dha} applies to a massless gas of
particles with Maxwell-Boltzmann distribution in a radiation dominated
epoch and it only approaches a stationary distribution at
asymptotically large times, while the present one applies generically
to a relativistic distribution in an inflationary regime and reaches
a stationary state in only a few e-folds.

\begin{figure}[!htb]
\begin{center}
  \includegraphics[width=0.6\textwidth]{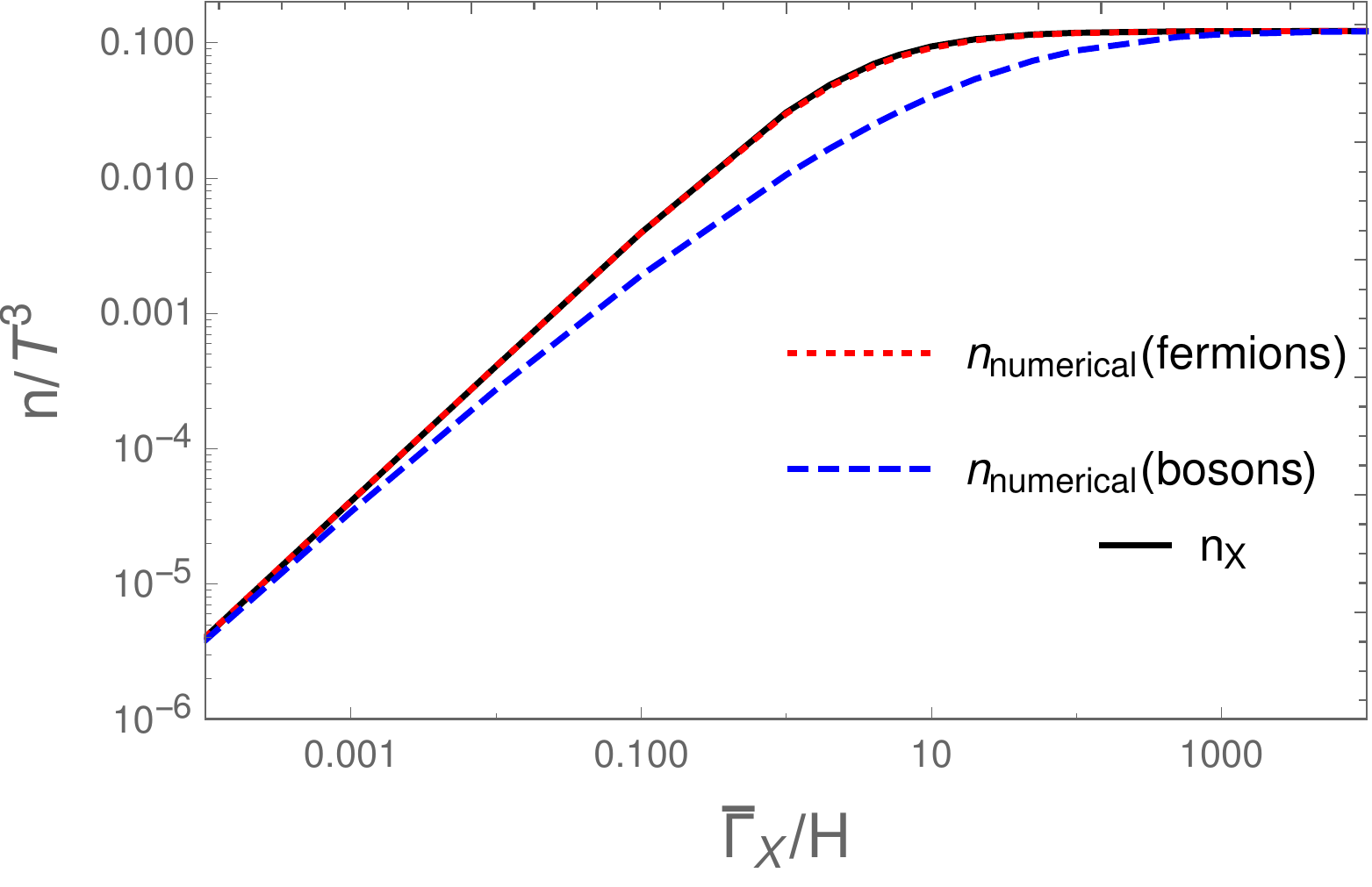}
\end{center}
\caption{The momentum integrated distributions for the cases of decay
  into bosons (dashed line) and decay into fermions (dotted line). The
  solid line is the result of the solution given by
  Eq.~(\ref{stationary_sol}).}
\label{fig4} 
\end{figure}

{}Finally, we can obtain explicitly the number density from our
results. We have integrated the numerically obtained distributions
over momenta and compared the results with the adiabatic solution for
the number density in Eq.~(\ref{stationary_sol}). These results are
shown in {}Fig.~\ref{fig4}. 

%%%%%%%%%%%%%%%%%%%%%%%%%%%%%%%%%%%%%%%%%%%%%%%%%%%%%%%%%%%%%%%%%%%%%%%
%%%%%%%%%%%%%%%%%%%%%%%%%%%%%%%%%%%%%%%%%%%%%%%%%%%%%%%%%%%%%%%%%%%%%%%%

We conclude that Eq.~(\ref{stationary_sol}) is a good approximation to
the numerical value of the number density, particularly for fermionic
decays, while again we observe some discrepancies in the case of
bosonic decays for intermediate values of $\bar\Gamma_X/H$. These only
occur for relativistic $X$ bosons and Eq.~(\ref{stationary_sol})
becomes a better approximation as $m_X$ increases. We note that, in
any case, the observed discrepancies correspond at most to
$\mathcal{O}(1)$ factors, so that $\gamma_X/(3+\gamma_X)n_X^{eq}$ is
generically a good approximation to the number density in the
adiabatic evolution regime.

Our generic analysis would not be complete without considering
scattering processes, which may also contribute to the collision term
in the Boltzmann equation. Although, for the same type of
interactions, scattering processes are typically suppressed compared
to decays since they involve larger powers of the couplings in the
perturbative regime, there may be cosmological settings where they
correspond to different types of interactions and may dominate the
collision term. The analysis of scattering processes is somewhat more
involved than for decays, since in general one cannot easily express
the collision term in terms of the number density, as we detail in
Appendix~\ref{AppA}. This is nevertheless possible in the limit of
small occupation numbers, $f_i\ll1$, i.e., in the absence of Bose
enhancement or {}Fermi degeneracy~\cite{Kolb:1990vq}. In this regime,
the Boltzmann equation can be written as
\begin{equation} \label{boltzmann_scatterings}
\dot{n}_X+3Hn_X\simeq  - \langle \sigma v\rangle\left(n_X^2-n_X^{eq\,
  2}\right)~,
\end{equation}
where $\langle \sigma v\rangle$ is the thermally averaged cross
section times velocity defined in Appendix~\ref{AppA}. In the
adiabatic limit, where both the latter and $H$ vary slowly, we obtain
the adiabatic solution for the number density
\begin{equation}
n_X\simeq {3\over2\gamma_X^{(s)}}\left(\sqrt{1+{4\over
    9}\gamma_X^{(s)\,2}}-1\right)n_X^{eq}~,
\end{equation}
where $\gamma_X^{(s)}= \langle \sigma v \rangle n_X^{eq}/H$ is the
ratio between the thermally averaged scattering rate and the Hubble
parameter. Although this may seem more  complicated than the adiabatic
solution for decays, we note that the suppression factor with respect
to the equilibrium number density becomes $\gamma_X^{(s)}/3$ for small
values of $\gamma_X^{(s)}$ and also tends to 1 in the limit
$\gamma_X\gg 1$, as in Eq.~(\ref{stationary_sol}). The cases where the
collision term is dominated by decays or by scattering processes thus
yield quite similar adiabatic solutions for the number density, at
least for small occupation numbers. 

%%%%%%%%%%%%%%%%%%%%%%%%%%%%%%%%%%%%%%%%%%%%%%%%%%%%%%%%%%%%%%%%%%%%%%%%%%%
%%%%%%%%%%%%%%%%%%%%%%%%%%%%%%%%%%%%%%%%%%%%%%%%%%%%%%%%%%%%%%%%%%%%%%%%%%%
%%%%%%%%%%%%%%%%%%%%%%%%%%%%%%%%%%%%%%%%%%%%%%%%%%%%%%%%%%%%%%%%%%%%%%%%%%%

\section{Adiabatic baryogenesis during warm inflation}
\label{sec3}

The fact that we have obtained new solutions to the Boltzmann equation
that are inherently out-of-equilibrium immediately suggests an
application: the production of a baryon asymmetry during
inflation. Let us then consider a generic model where the $X$
particles (bosons or fermions) decay violating $B$ (or $L$ or $B-L$) and
C/CP. In particular, let us consider a simple case with two possible
decay channels (as discussed e.g.~in Ref.~\cite{Cline:2006ts}),
\begin{eqnarray}
X\rightarrow B~, \qquad X\rightarrow Y~,
\end{eqnarray}
where the final state $B$ carries a baryon number $b>0$ and $Y$ has no
baryonic charge. In practical applications these will typically
correspond to 2-body decays, although the number of particles in the
final state can be arbitrary. $B$-violation then occurs because there is
no consistent assignment of a baryonic charge to the $X$ particle. We
also have the conjugate decays $\bar{X}\rightarrow \bar{B}$ and
$\bar{X}\rightarrow \bar{Y}$ with opposite baryonic charges, and C/CP
violation implies that the partial decay widths satisfy the relations
\begin{eqnarray}
\Gamma(X\rightarrow B)\!\!&=&\!\!{1\over2}(1+\epsilon)\Gamma_X~,\quad
\Gamma(X\rightarrow
Y)={1\over2}(1-\epsilon)\Gamma_X~,\nonumber\\ \Gamma(\bar{X}\rightarrow
\bar{B})\!\!&=&\!\!{1\over2}(1+\bar{\epsilon})\Gamma_X~,\quad
\Gamma(\bar{X}\rightarrow
\bar{Y})={1\over2}(1-\bar{\epsilon})\Gamma_X~,
\end{eqnarray}
where $\epsilon\neq \bar{\epsilon}$ yield the amount of C/CP
violation. Note that the total decay widths of the $X$ particle and of
its anti-particle are equal, as required by CPT invariance,
\begin{eqnarray}
\Gamma(X\rightarrow B)+\Gamma(X\rightarrow
Y)=\Gamma(\bar{X}\rightarrow \bar{B})+\Gamma(\bar{X}\rightarrow
\bar{Y})~.
\end{eqnarray}
Both $X$ and $\bar{X}$ will then obey the same Boltzmann
equation~(\ref{Boltzmann}) when the particles in the $B, Y$ final
states are in equilibrium (which we assume to be maintained by other
interactions), and evolve towards the adiabatic
solution~(\ref{stationary_sol}), which will be the same for both $n_X$
and $n_{\bar{X}}$. The full adiabatic solution~(\ref{adiabatic_sol})
may be different for both $n_X$ and $n_{\bar{X}}$ if the initial
values are distinct, but any discrepancies are quickly erased by
expansion.

Now, the baryon number density, i.e., the difference between the
number density of baryons and that of anti-baryons, evolves according
to the Boltzmann equation given by
\begin{eqnarray} \label{Boltzmann_baryon}
n_B'+3 n_B=b\left[{\gamma_X\over
    2}(1+\epsilon)(n_X-n_X^{eq})-{\gamma_X\over
    2}(1+\bar{\epsilon})(n_{\bar{X}}-n_{\bar{X}}^{eq})\right]~.
\end{eqnarray}
Let us consider the case where $n_X(0)=n_{\bar{X}}(0)$, such that
\begin{eqnarray} \label{Boltzmann_baryon_1}
n_B'+3 n_B= - b{\gamma_X\over 2}\Delta\epsilon(n_X-n_X^{eq})~,
\end{eqnarray}
where $\Delta\epsilon =\bar{\epsilon}-\epsilon$ corresponds to the
amount of CP violation. Using the solution given by
Eq.~(\ref{adiabatic_sol}), this yields the adiabatic solution for the
baryon number density,
\begin{eqnarray} \label{Boltzmann_baryon_sol}
n_B(N_e)&\simeq & {b\over
  2}\Delta\epsilon\left[\left(1-e^{-(3+\gamma_X)N_e}\right){\gamma_X\over
    3+\gamma_X}n_X^{eq}+n_X(0)e^{-3N_e}
  \left(e^{-\gamma_XN_e}-1\right)\right]~,
\end{eqnarray}
which approaches (exponentially fast) the quasi-stationary solution,
\begin{eqnarray} \label{baryon_stationary}
n_B\simeq {b\over 2}\Delta\epsilon{\gamma_X\over 3+\gamma_X}n_X^{eq}~.
\end{eqnarray}
Hence, we see that a constant baryon asymmetry is produced during warm
inflation (or in fact during any analogous period of quasi-adiabatic
temperature and Hubble rate evolution), for any value of
$\gamma_X= \bar{\Gamma}_X/H$. Interestingly, $n_B\rightarrow b\Delta\epsilon
n_X^{eq}/2$ for $\gamma_X\rightarrow \infty$, the limit for which the
$X$ particles are in equilibrium. This may seem to contradict
Sakharov's conditions for baryogenesis~\cite{Sakharov:1967dj}, but it is simply associated with the
fact that we are taking decays and inverse decays as the main
processes responsible for driving the $X$ particles towards
equilibrium. In this case, to get closer to equilibrium, we need to
increase $\gamma_X$, which also increases the rate of production of
baryon number, in such a way that we obtain a finite baryon number
density for arbitrarily large $\gamma_X$.  Note that, in the large $\gamma_X$ limit, from the solution
given by Eq.~(\ref{adiabatic_sol}), we have that $n_X - n_X^{eq} \to -3 n_X^{eq}/\gamma_X$.
Thus, the baryon source term on the right-hand-side of Eq.~(\ref{Boltzmann_baryon_1}) tends to a finite value 
$3 b \Delta\epsilon n_X^{eq}/2$  in the limit $\gamma_X \to +\infty$. However, note that in any
physical setting $\gamma_X$ is not at the limiting value
but rather is finite, which means the $X$ particles are
always out-of-equilibrium.  Nevertheless this analysis 
demonstrates that this parameter
can be arbitrarily large and still produce a significant baryon
asymmetry.

The baryon asymmetry produced during warm inflation can set the final
cosmological asymmetry if the source term in the baryon number density
equation becomes suppressed after the slow-roll regime and throughout the subsequent the cosmic history. This is, of
course, model-dependent, but we can envisage scenarios where some
other processes, such as e.g.~scatterings, are more suppressed
than decays during warm inflation, but become the dominant processes
once the slow-roll period is over and radiation becomes the dominant
component. In this case $X$ will be driven towards equilibrium after
inflation more quickly than through decays and inverse decays and the
baryon source will quickly shut down. If it remains in equilibrium
until it is sufficiently non-relativistic, there should be no
significant sources of baryon number at late times that could
substantially modify the asymmetry produced during
inflation\footnote{Potentially electroweak
  sphalerons~\cite{Klinkhamer:1984di} may convert a lepton asymmetry
  into a baryon asymmetry in a leptogenesis
  scenario~\cite{Fukugita:1986hr}.}.

The smallness of the observed cosmological baryon-to-entropy ratio
$n_B/s$ can have different sources in this scenario: (i) CP-violation
may be small, $\Delta\epsilon\ll 1$, (ii) the $X$ particles may be far
from equilibrium, $\gamma_X\ll 1$ or (iii) $X$ particles may be
non-relativistic during warm inflation, $n_X^{eq}/s \ll 1$. Any
combination of these may thus easily explain why $n_B/s\sim 10^{-10}$
(see, e.g., Ref.~\cite{etaBBN} for BBN constraints on this ratio).

Let us consider in more detail the particular case where the $X$
particles are relativistic during inflation, $m_X\ll T$, as well as
fully out-of-equilibrium, $\gamma_X\ll 1$. In this case, the
baryon-to-entropy ratio is given by
\begin{eqnarray} \label{baryon_entropy_stationary}
{n_B\over s}\simeq {45\zeta(3)\over 4\pi^4}b\Delta\epsilon{g_X\over
  g_*}{\bar{\Gamma}_X\over H}~,
\end{eqnarray}
where $g_X$ is the number of degrees of freedom in $X$, to which each
bosonic (fermionic) degree of freedom contributes by a factor 1 (3/4),
and $g_*$ is the effective total number of relativistic degrees of
freedom in the thermal bath. The smallness of the observed
baryon-to-entropy ratio may in this case be due to a small amount of
CP violation in $\Delta\epsilon$ and/or large deviations from thermal
equilibrium, $\gamma_X\ll 1$, during inflation, or a combination of
both these factors, with an additional suppression by $g_X/g_*$. This
will of course be a model-dependent issue that is not pursued
further here.
We note that the high temperatures typically attained
during warm inflation (see, e.g., Refs.~\cite{Bartrum:2013fia,
  Bastero-Gil:2016qru}) suggest possible implementations of this
mechanism within grand unified theories~\cite{Nanopoulos:1979gx}.

%%%%%%%%%%%%%%%%%%%%%%%%%%%%%%%%%%%%%%%%%%%%%%%%%%%%%%%%%%%%%
%%%%%%%%%%%%%%%%%%%%%%%%%%%%%%%%%%%%%%%%%%%%%%%%%%%%%%%%%%%%%
%%%%%%%%%%%%%%%%%%%%%%%%%%%%%%%%%%%%%%%%%%%%%%%%%%%%%%%%%%%%%

\section{Generation of Baryon Isocurvature Perturbations}
\label{sec4} 

The interesting aspect that we would like to discuss in more detail is
the fact that $n_B/s$ depends on the ratio $\bar{\Gamma}_X/H$, which, albeit
nearly constant, exhibits a small variation during inflation and,
moreover, is associated with the inflationary dynamics. As such, this
ratio will necessarily acquire fluctuations on superhorizon scales due
to fluctuations in the inflaton field. This implies that the latter
will induce both adiabatic curvature fluctuations and baryon
isocurvature fluctuations, the latter corresponding to relative
fluctuations in the baryon and photon fluids, which would be absent if
the baryon asymmetry were not generated during inflation. This is
similar to the {\it warm baryogenesis} scenario proposed in
Ref.~\cite{BasteroGil:2011cx}, where the baryon asymmetry is directly
sourced by the dissipative processes that sustain the thermal bath in
warm inflation, but the spectrum of isocurvature modes may be
different. This yields the interesting possibility of looking for
baryon isocurvature modes with CMB and LSS observations to assess
whether the observed baryon asymmetry was or not produced during
inflation, but also whether its generation was directly linked with
dissipative dynamics. Note that other baryogenesis
models~\cite{isocurvature}, such as Affleck-Dine baryogenesis, may
also lead to baryon isocurvature modes in the primordial spectrum, but
these are in this case uncorrelated with the main adiabatic curvature
component, since they are associated with distinct fields. The degree
of correlation between isocurvature and adiabatic curvature modes may
thus be used to test the warm inflation paradigm itself and isolate a
particular mechanism for baryogenesis.

Let us then analyze in more detail the spectrum of baryon isocurvature
perturbations produced in the present model. Since $n_B/s$ will always
depend on $\gamma_X$ in the adiabatic dynamics under consideration,
baryon isocurvature modes will always be generated, independently of
the value of $\gamma_X$ or whether $X$ particles are relativistic
during inflation, but let us focus, for concreteness, on the
weakly-coupled high-temperature case leading to
Eq.~(\ref{baryon_entropy_stationary}). At high temperature, we
typically have that $\bar{\Gamma}_X\propto T$, such that $n_B/s\propto T/H$.
Baryon isocurvature modes are characterized by the
quantity~\cite{Lyth:2002my}
\begin{eqnarray} \label{isocurvature_1}
S_B= {\delta\rho_B\over \rho_B}-{3\over4}{\delta\rho_R\over \rho_R} =
{\delta(n_B/s)\over n_B/s} ={\delta (T/H)\over T/H}
\end{eqnarray}
evaluated when the relevant CMB scales become superhorizon during
inflation.  The subscript `R' used in Eq.~(\ref{isocurvature_1}) and
in the quantities below refers to radiation. We thus need to determine
how the fluctuations in the ratio $T/H$ are related to inflaton
fluctuations. The dynamics of warm inflation is dictated by the
coupled inflaton and radiation equations, which for the homogeneous
background components are given by
\begin{eqnarray} \label{warm_eq_phi}
\ddot\phi + (3H+\Upsilon)\dot\phi + V'(\phi)=0~,
\end{eqnarray}
\begin{eqnarray} \label{warm_eq_rad}
\dot\rho_R + 4H\rho_R = \Upsilon\dot\phi^2~,
\end{eqnarray}
where $\Upsilon$ is the dissipation coefficient. In the slow-roll
regime, valid when the slow-roll parameters $\epsilon_\phi,
|\eta_\phi| \ll 1+Q$, where $Q=\Upsilon/3H$,
$\epsilon_\phi=(M_P^2/2)(V'/V)^2$ and $\eta_\phi= M_P^2 V''/V$, these
become
\begin{eqnarray} \label{warm_eqs_sr}
\dot\phi \simeq -{V'(\phi)\over 3H(1+Q)}~, \qquad \rho_R\simeq {3\over
  4}Q \dot\phi^2~.
\end{eqnarray}
Combining both equations for $\rho_R= C_R T^4$, where
$C_R=(\pi^2/30)g_*$, we obtain after some algebra, that
\begin{eqnarray} \label{warm_eqs_3}
\left({T\over H}\right)^4\simeq {3\over 2}C_R^{-1}\left({M_P\over
  H}\right)^2 {Q\over (1+Q)^2}\epsilon_\phi~,
\end{eqnarray}
where one can see that a warm thermal bath, i.e., ~$T\gtrsim H$, can
easily be attained for $H\ll M_P$ in the slow-roll regime even if the
dissipative ratio $Q$ is not very large. Considering perturbations in
the above equation, we obtain after a straightforward calculation,
\begin{eqnarray} 
S_B= {\delta(T/H)\over T/H} \simeq {1\over
  4}\left({6\epsilon_\phi-2\eta_\phi\over 1+Q}+ {1-Q\over 1+Q}
{Q'\over Q}\right)\mathcal{R}~,
\end{eqnarray}
where $\mathcal{R}\simeq (H/\dot\phi) \delta\phi$ is the
gauge-invariant comoving curvature perturbation (written in the
$\Psi=0$ gauge~\cite{lyth2009primordial}) and primes denote
derivatives with respect to the number of e-folds of inflation, $dN_e=
Hdt$. The dynamical quantity $Q'/Q$ depends on the form of the
dissipation coefficient $\Upsilon$, but it is in general a linear
combination of the slow-roll parameters divided by a linear polynomial
in $Q$ (see, e.g., Refs.~\cite{BasteroGil:2009ec, Bartrum:2013fia,
  Bastero-Gil:2016qru}). This implies that typically
$S_B/\mathcal{R}\sim \mathcal{O}(n_s-1)\sim \mathcal{O}(N_e^{-1})$. 

The {\it Warm Little Inflaton}  (WLI) scenario of
Ref.~\cite{Bastero-Gil:2016qru}, where the inflaton interacts with
relativistic fermion fields and $\Upsilon\propto T$, constitutes the
simplest and most appealing particle physics realization of
warm inflation. In this case, we have $Q'/Q=
(6\epsilon-2\eta_\phi)/(3+5Q)$. {}For thermal inflaton fluctuations
and weak dissipation at horizon-crossing, $Q_*\ll1$, we have for the
scalar spectral index $n_s-1= 2(2\eta_{\phi_*}-6\epsilon_{\phi_*})/3$,
yielding %
\begin{eqnarray} 
S_B^{\rm WLI}\simeq  \left({1-n_s\over 2}\right)\mathcal{R}\simeq
10^{-2}\mathcal{R}~.
\end{eqnarray}
Although the exact relation between $S_B$ and $\mathcal{R}$ is
model-dependent, we expect these quantities to be in general
proportional with a proportionality constant of this magnitude as
argued above. We note that the effects of baryon and cold dark matter
isocurvature modes (CDI) on the CMB spectrum are indistinguishable,
although, e.g.,~the trispectrum may in principle distinguish between
them~\cite{Grin:2013uya}. As such, the effective contribution to the
cold dark matter isocurvature spectrum from the baryon modes above is
given by   %
\begin{eqnarray} 
P_{CDI} = \left({\Omega_B\over \Omega_c}\right)^2 \left({S_B\over
  \mathcal{R}}\right)^2 P_\mathcal{R}~.
\end{eqnarray}
The Planck analysis of CDI modes in the CMB
spectrum~\cite{Ade:2015lrj} uses the variable %
\begin{eqnarray}
  \beta_{ISO}= {P_{CDI}\over P_\mathcal{R}+P_{CDI}} \,,
%  \sim 10^{-5}~,
\end{eqnarray}
and for the WLI model this gives $\beta_{ISO} \sim 10^{-5}$ 
when $n_s\simeq 0.96-0.97$, which, as argued above, should also give the
generic magnitude of the effect. This is still well below the
state-of-the-art constraints set by the Planck collaboration, yielding
$\beta_{ISO}\lesssim 10^{-2}$ for generic CDI models and using
different data sets~\cite{Ade:2015lrj}. Particular models of CDI modes
can be constrained by an additional order of magnitude, but the
predictions of our baryogenesis mechanism are still compatible with
the Planck results and may be tested in the future with increased
precision measurements.

The proportionality between $S_B$ and $\mathcal{R}$ shows that these
quantities are naturally correlated, since both adiabatic and
isocurvature modes are generated by thermal inflaton fluctuations.
Their correlation is, however, scale-dependent, since the baryon
isocurvature spectral index differs from the adiabatic spectral index,
\begin{eqnarray} 
1-n_{I}&\simeq &{d\ln P_{CDI}\over dN_e} =  {d\ln P_\mathcal{R}\over
  dN_e} -{2n_s'\over 1-n_s} \simeq
(1-n_s)\left[1-{2n_s'\over (1-n_s)^2}\right]~.
\end{eqnarray}
The correction due to the running of the adiabatic spectral index can
be significant, since in most models $n_s' \sim
\mathcal{O}((1-n_s)^2)$. {}For instance, for thermal inflaton
fluctuations and weak dissipation at horizon-crossing, $Q_*\ll 1$, %
\begin{eqnarray} 
n_s' \simeq {2\over 3}\left(-2\xi_{\phi_*}^2
+16\epsilon_{\phi_*}\eta_{\phi_*}-24\epsilon_{\phi_*}^2\right)~,
\end{eqnarray}
where $\xi_\phi^2= M_P^4 V''' V'/V^2$. If we consider a quartic
chaotic inflaton potential, $V(\phi)=\lambda \phi^4$, which yields
predictions for both $n_s$ and the tensor-to-scalar ratio $r$ in
excellent agreement with the Planck results within the WLI scenario
for warm inflation~\cite{Bastero-Gil:2016qru}, one finds
$n_s'=-(1-n_s)^2/2$, such that $(1-n_I)\simeq 2(1-n_s)$, yielding
$n_I\sim 0.92-0.94$ for $n_s=0.96-0.97$. This implies, in particular,
that the Planck constraints for CDI modes fully correlated with the
main adiabatic component with $n_I=n_s$ do not apply in the present
scenario.

Note that for larger values of $\gamma_X$, the baryon-to-entropy ratio
becomes less dependent on the latter, thus suppressing the associated
baryon isocurvature modes. Also, when $X$ is
non-relativistic, $n_B/s$ exhibits an additional dependence on the
ratio $m_X/T$ that must be taken into account. This may potentially
enhance $S_B$, depending on the particular model of warm inflation
considered, although relative fluctuations in the ratio above are also
typically proportional to combinations of slow-roll parameters, and
hence necessarily $\mathcal{O}(n_s-1)$.

%%%%%%%%%%%%%%%%%%%%%%%%%%%%%%%%%%%%%%%%%%%%%%%%%%%%%%%%%%%%%%%%%%
\section{Summary and future prospects}
\label{sec5}

In this work we have shown that particle number densities during a
period of warm inflation can follow out-of-equilibrium adiabatic
solutions to the Boltzmann equation and which are suppressed relative
to the equilibrium value by a factor $\gamma_X/(3+\gamma_X)$, where
$\gamma_X=\bar\Gamma_X/H$ is the ratio between the thermally averaged
decay rate of the particle species $X$ of interest and the inflationary Hubble rate, obtaining a similar result for the case
where scattering processes yield the dominant interactions. We have
also shown numerically that the corresponding phase
space-distributions tend to a stationary configuration with a modified
equilibrium distribution, given by  Eq.~(\ref{fit}), with essentially
the above amplitude suppression (up to some distortion due to the
momentum-dependence of the decay width) and a slightly smaller
effective temperature. Such adiabatic solutions are achieved after a
small number of e-folds that naturally decreases when $\gamma_X$
increases.

This shows that particles can remain out-of-equilibrium throughout
warm inflation, with small but not necessarily negligible number
densities. To illustrate the impact of this result, we have shown that
the observed cosmological baryon asymmetry could be produced by the
out-of-equilibrium decay of a generic $X$ particle interacting with
the inflationary thermal bath, violating baryon number and C/CP. The
smallness of the resulting baryon asymmetry can in this case be a
consequence of the small value of $\gamma_X$ in combination with a
small amount of CP violation, and also of the Boltzmann suppression in
the case of non-relativistic particles. An interesting feature of this
generic scenario is the generation of superhorizon baryon isocurvature
modes, correlated with the main adiabatic curvature perturbations. The
spectrum of such modes is model-dependent but we have shown that
generically the predicted amplitude is below the current constraints
on (effective) cold dark matter isocurvature perturbations by the
Planck collaboration. Evidence for such modes could, in the future,
constitute a smoking gun for the production of a baryon asymmetry
during inflation and, in fact, for a warm inflation scenario, and the
detailed properties of the spectrum may help to distinguish the
present scenario from the warm baryogenesis mechanism, where a baryon
asymmetry is produced directly by the dissipative effects that sustain
the thermal bath during inflation.

As for the warm baryogenesis mechanism, the present scenario may also
be generalized to an asymmetry in other particles carrying different
charges, possibly producing an asymmetry in the dark matter
sector. The resulting CDI power spectrum has an amplitude larger than
the baryonic modes by a factor $(\Omega_c/\Omega_B)^2\sim 30$
\cite{Ade:2015xua}, which may thus be more easily probed.

Our results may have a significant impact in other aspects of the
inflationary dynamics. For instance, it is often assumed that the
particles interacting directly with the inflaton and dissipating its
energy are in equilibrium with the overall thermal bath. This
is required for consistently using equilibrium phase-space
distribution functions to compute the associated dissipation
coefficients (see, e.g., Refs.~\cite{BasteroGil:2010pb,
  BasteroGil:2012cm}). Typically this requires such particles to decay
faster than the Hubble rate, posing constraints on the coupling
constants and particle masses considered that could be relaxed if
out-of-equilibrium distribution functions are known. Our results can
thus potentially be employed to this effect, a possibility that we
plan to investigate in detail in future work.

Another important aspect of warm inflation where the results obtained
in this work should be of relevance is the primordial spectrum of
curvature perturbations, since the latter depends on the phase
space distribution of inflaton fluctuations. In particular, for weak
dissipation at horizon-crossing, predictions for $n_s$ and $r$ differ
significantly for the limiting cases where inflaton fluctuations are
in a vacuum or in a thermal state (i.e., in equilibrium with the
overall thermal bath)~\cite{Bartrum:2013fia, Bastero-Gil:2016qru}. Although for strong dissipation this issue
becomes less relevant, since dissipation becomes the dominant source
of inflaton fluctuations, agreement with observations in most
scenarios considered so far typically favours the $Q_*\lesssim 1$
regime~\cite{Bartrum:2013fia, Bastero-Gil:2016qru, Benetti:2016jhf, Bastero-Gil:2017wwl, Arya:2017zlb}. The dissipative dynamics
itself is not sufficient to determine the state of inflaton
fluctuations, since other processes in the thermal bath can be
responsible for a substantial creation and annihilation of inflaton
particles. Since the inflaton's direct interactions with other
particles cannot typically be very strong, it is unlikely that full
thermal equilibrium of inflaton particles is achieved in general.
Nevertheless, production of inflaton
particles may play a substantial role. The adiabatic
solutions obtained in this work could then be used to infer the
inflaton phase-space distribution at horizon-crossing in different
models, eliminating the uncertainty in observational predictions.

Adiabatic solutions to the Boltzmann equation require both the ratio
$\Gamma_X/H$ and the equilibrium distribution $n_X^{eq}$ to vary
slowly compared to the expansion rate, thus requiring both $H$ and $T$
to remain nearly constant. This naturally makes warm inflation
the type of dynamics to which such solutions can be applied.

We may envisage, however, other
cosmological scenarios where, in addition to the early inflationary
period in which the observable CMB scales became superhorizon, there are
other (shorter) periods of inflation where particle production can
sustain the temperature of the cosmic thermal bath. This may be, for
instance, the case of second order or crossover cosmological phase
transitions, where a scalar field rolls to a new minimum once the temperature drops below a critical value. It has
been shown in Ref.~\cite{Bartrum:2014fla} that dissipative
friction can make the field's vacuum energy dominate over the
radiation energy density, and in fact prevent the latter from
redshifting due to expansion. This may e.g.~dilute unwanted thermal relics produced during or after the first period of warm inflation. Our solutions may, thus, also describe the evolution of the
number density of different particle species during such periods,
which may have a significant impact on their present abundances. One
can consequently also envisage baryogenesis scenarios along the lines
proposed above during these shorter inflationary periods, although
these may not easily be tested if the associated isocurvature
perturbations are generated at too small scales.
 
 We should not exclude other applications of our solutions where both particle production and an
expanding environment are involved and have analogous adiabatic conditions
as the ones we considered here, e.g., possibly in the
quark-gluon plasma formation and subsequent hadronization process under
study with heavy-ion collisions experiments. 

In summary, the novel adiabatic solutions to the Boltzmann equation
found in this work can have a significant impact in cosmology and shed
a new light on several of its presently open questions.

%%%%%%%%%%%%%%%%%%%%%%%%%%%%%%%%%%%%%%%%%%%%%%%%%%%%%%%%%%%%%%%%%%%%%%%%
\acknowledgments{
A.\,B.~is supported by STFC. M.\,B.\,-G.~is partially supported by
``Junta de Andaluc\'ia'' (FQM101) and MINECO (Grant
No. ~FIS1016-78198-P). R.\,O.\,R.~is partially supported by  Conselho Nacional
de Desenvolvimento Cient\'{\i}fico e Tecnol\'ogico - CNPq (Grant
No.~303377/ 2013-5) and Funda\c{c}\~ao Carlos Chagas Filho de Amparo
\`a Pesquisa do Estado do Rio de Janeiro - FAPERJ (Grant
No.~E-26/201.424/2014). J.\,G.\,R. is supported by the FCT
Investigator Grant No.~IF/01597/2015 and partially by the
H2020-MSCA-RISE-2015 Grant No. StronGrHEP-690904 and by the CIDMA
Project No.~UID/MAT/04106/2013. }

%%%%%%%%%%%%%%%%%%%%%%%%%%%%%%%%%%%%%%%%%%%%%%%%%%%%%%%%%%%%%%%%%%%%%%%%
\appendix

\section{Boltzmann equation for scattering processes}
\label{AppA}

Let us consider, for concreteness, the case where $X$ bosons can
annihilate into $Y$ bosons in the thermal bath, $XX\leftrightarrow
YY$, although our analysis can be easily generalized to different
types of particles and processes such as $XY\leftrightarrow
XY$. Labeling the $X$ particles as $(1,2)$ and the $Y$ particles as
$(3,4)$, the collision term in the Boltzmann equation for the $X$
number density, Eq.~(\ref{boltzmann_n}) is given by %
\begin{eqnarray}  \label{collision_scattering}
C&=&\!\!-\!\!\!\int\! d\Pi_1 d\Pi_2 d\Pi_3
d\Pi_4(2\pi)^4\delta^4\left(p_1+p_2-p_3-p_4\right)|\mathcal{M}|^2
\nonumber\\ &\times
&\left[f_1f_2(1+f_3^{eq})(1+f_4^{eq})-f_3^{eq}f_4^{eq}(1+f_1+f_2)\right]
\end{eqnarray}
where $\mathcal{M}$ is the scattering matrix element,
and we take the $Y$ particles as part of the thermal bath.
Using conservation of energy and the form of the equilibrium 
distributions, we can show that: %
\begin{eqnarray} 
{1+f_3^{eq}+f_4^{eq}\over f_3^{eq} f_4^{eq}}= {1+f_1^{eq}+f_2^{eq}\over f_1^{eq}
  f_2^{eq}}~,
\end{eqnarray}
such that we may write the term in square brackets in
Eq.~(\ref{collision_scattering}), after some algebra, as %
\begin{eqnarray} 
\!\!\!\!\!\!{f_3f_4\over
  f_1^{eq}f_2^{eq}}\!\!\left[f_1f_2-f_1^{eq}f_2^{eq}+\sum_{i\neq
    j=1,2}\!\! f_if_i^{eq}(f_j-f_j^{eq})\right]~,
\end{eqnarray}
which clearly vanishes in equilibrium. The collision term can then be
written in the form %
\begin{eqnarray}  \label{collision_scattering_2}
C=-\int {d^3p_1\over (2\pi)^3}{d^3p_2\over (2\pi)^3}\!\!
\bigg[f_1f_2-f_1^{eq}f_2^{eq}+\!\!\!\!\!\sum_{i\neq j=1,2}\!\!\!\!\!
  f_if_i^{eq}(f_j-f_j^{eq})\bigg] \sigma v,
\end{eqnarray}
where the cross section times velocity factor is given by %
\begin{eqnarray}  \label{sigma_v}
\sigma v ={(f_1^{eq}f_2^{eq})^{-1}\over 4E_1E_2}\!\!\!\int
\!\!d\Pi_3d\Pi_4|\mathcal{M}|^2
(2\pi)^4\!\delta^4\left(p_1+p_2-p_3-p_4\right)f_3^{eq}f_4^{eq}.
\end{eqnarray}
The cubic terms in the phase-space distribution functions
$f_if_i^{eq}(f_j-f_j^{eq})$ prevent writing the collision term in
terms of the number density in a simple form, but may be discarded
when $f_i\ll 1$, in which case $f_i\simeq e^{-E_i/T}$ (discarding
chemical potentials for simplicity). In this case we can write the
collision term as %
\begin{eqnarray}  \label{collision_scattering_3}
C\simeq - \langle \sigma v\rangle (n_X^2-n_X^{eq\, 2})~,
\end{eqnarray}
where we approximated the energy dependent cross section times
velocity of the scattering process by its thermal average \cite{Kolb:1990vq}, as done for
the case of decays, %
\begin{eqnarray}  \label{collision_scattering_4}
 \langle \sigma v\rangle = {1\over n_X^{eq\, 2}} \!\!\int
 \!\!{d^3p_1\over (2\pi)^3}{d^3p_2\over (2\pi)^3} \sigma v
 f_1^{eq}f_2^{eq}~.
 \end{eqnarray}
This then yields Eq.~(\ref{boltzmann_scatterings}) for the Boltzmann
equation when scattering processes dominate the collision term.

%%%%%%%%%%%%%%%%%%%%%%%%%%%%%%%%%%%%%%%%%%%%%%%%%%%%%%%%%%%%%%%%%%%%%%%%
%\newpage

\end{document}